\documentclass[sigconf,nonacm]{acmart}
\usepackage{cleveref}

\AtBeginDocument{%
  }

\newcommand{\sd}[2]{$\underset{(#2)}{#1}$}

\begin{document}

\title{Revisiting Elliptic++ for Bitcoin Forensics: Feature Reconstruction and Leakage-Resistant Evaluation}

\author{Runlin Liang}
\authornote{Corresponding author.}
\affiliation{%
  \institution{University of Amsterdam}
  \city{Amsterdam}
  \country{Netherlands}\\
  \institution{Tinbergen Institute}
  \city{Amsterdam}
  \country{Netherlands}
}
\email{r.liang@uva.nl}

\author{Simon Trimborn}
\affiliation{%
  \institution{University of Amsterdam}
  \city{Amsterdam}
  \country{Netherlands}\\
  \institution{Tinbergen Institute}
  \city{Amsterdam}
  \country{Netherlands}
  }
\email{s.trimborn@uva.nl}

\author{Noud van Giersbergen}
\affiliation{%
  \institution{University of Amsterdam}
  \city{Amsterdam}
  \country{Netherlands}
  }
\email{N.P.A.vanGiersbergen@uva.nl}

\begin{abstract}
Elliptic and its address-level extension Elliptic++ are widely used benchmark datasets in Bitcoin forensics, cited by more than 1{,}300 academic studies according to Google Scholar. We identify future information leakage in Elliptic++. Its positional data split does not respect temporal order, and the exact definitions of its address-level features remain undocumented. We revisit how the 55 Elliptic++ address-level features are computed and propose a leakage-resistant pipeline. First, the exact feature definitions are recovered by linking Elliptic transactions to public ledger data. Second, we compute the features with a temporal split based on each address's first appearance, using only information available at the time of prediction. Once temporal ordering is enforced, the random forest (RF)'s strong predictive performance largely disappears: its F1 score on the illicit class falls from $0.846$ under the original leaky evaluation to just $0.291$ under our leakage-resistant pipeline. Furthermore, RF no longer outperforms a simple logistic regression baseline. We provide the exact feature definitions and executable code to reproduce the leakage-resistant features, so that future Bitcoin forensic models can be evaluated based on genuine predictive capability rather than temporal data leakage.
\end{abstract}



\maketitle

\section{Introduction}
\label{sec:intro}
The pseudonymous nature of cryptocurrency transactions makes them attractive for illicit use. In 2025, the volume of cryptocurrency involved in identified illicit activities was estimated to be at least \$154 billion, more than double the \$57.2 billion recorded in 2024 \citep{chainalysis2026}. On top of pseudonymity, a public and decentralised ledger further encourages illicit use by removing the need for a trusted intermediary among transacting parties, though it simultaneously makes that use traceable. Publicly accessible transaction data allow forensic analysts to reconstruct flows of illicit funds and study criminal characteristics \citep{Meiklejohn2013, Christin2013, Huang2018, Trimborn2024}. Among all cryptocurrencies, Bitcoin has caught particular attention from academic researchers and empirical analysts, given its dominating market share. \citet{Foleyetal2019} estimate that approximately a quarter of Bitcoin users and half of its transactions were associated with illegal activity between 2009 and 2017. They also show that this illegal share declined over the period as mainstream adoption grew.

In practice, forensic analysts face far more candidate transactions and addresses than can be examined manually, creating an operational need for detection tools that flag suspicious activities. Recent literature treats this forensic detection as a supervised classification task, releasing labelled datasets of Bitcoin transactions or addresses, and benchmarking detection tools and algorithms on them. The Elliptic dataset \citep{WeberEtAl2019}, at its release the largest publicly available labelled Bitcoin transaction dataset, was made public jointly by the MIT-IBM Watson AI Lab and Elliptic, a blockchain intelligence firm that partners with UK government agencies\footnote{Elliptic announced on its webpage (\url{https://www.elliptic.co/media-center/elliptic-partners-with-uk-agencies-to-tackle-crypto-enabled-sanctions-evasion}) that it participated in a cross-government operation led by the Office of Financial Sanctions Implementation to identify and disrupt crypto crimes, together with several other UK agencies including National Crime Agency, Metropolitan Police Service, His Majesty's Revenue and Customs, Financial Conduct Authority, City of London Police, and Regional Organised Crime Units.}. Its labels are produced by Elliptic's investigators, who attribute transactions to known licit and illicit entities, such as exchanges, ransomware operators, and scams, through a heuristics-based reasoning process \citep{WeberEtAl2019}. With this provenance, the Elliptic dataset is commonly used as ground truth in the literature. Subsequent datasets have extended the labels to the address level (Elliptic++) \citep{ElmougyLiu2023} and to subgraphs of a much larger transaction graph (Elliptic2) \citep{BelleiEtAl2024}. 

On this family of Elliptic datasets, the reported detection performance is generally high, yet recent re-evaluations show that such performance advantages need not reflect out-of-sample detection, due to information leakage \citep{Maganti2026, KhaleghpourMcKinney2026, Safar2026}. A forensic detector must flag entities that appear after it is trained, so its evaluation must respect that order. Such inflated performance is also reported for malware detection in the security literature \citep{Pendlebury2019, Arp2022}. 

There are two main channels through which the leakage can enter: the split and the feature channels. The first one is a randomised split disregarding time, in which training and test samples can come from the same periods \citep{Pendlebury2019, Arp2022, Kapoor2023}. The second one is a problematic feature computation process, in which features are computed with information from the test period and consequently violate the legitimacy rule of \citet{Kaufman2012}\footnote{A feature is legitimate only if its value would be observable when the prediction is made: information observed at time $t_x$ may enter a prediction of a target realised at time $t_y$ only if $t_x < t_y$ \citep{Kaufman2012}.}. Elliptic's transaction-level split follows time, but \citet{Safar2026} reverse-engineered its anonymised transaction-level features and found that most of the nominally local features are aggregates of the transacting parties (addresses), computed with information that includes the test period. These transaction-level features are therefore leaky, through the feature channel. Moreover, because these address-aggregates are identical for every transaction that involves the same address, training and test transactions are linked through shared addresses, a further channel identified by \citet{Safar2026}. They argue that because the transactions are labelled by association between transactions' input addresses and the known entities controlling them, a classifier can then inflate its performance by recognising those addresses rather than by detecting illicit activity.

Both the Elliptic and Elliptic++ datasets are widely used in Bitcoin forensics research. As of 13 October 2026, Google Scholar lists 1{,}154 citations of \citet{WeberEtAl2019} and 205 of \citet{ElmougyLiu2023}. The leakage problem may therefore affect a large body of published and ongoing work, which calls for a solution. In Elliptic, transactions are labelled as illicit, whereas in Elliptic++, labels are assigned to addresses, which may serve as inputs to many transactions. For transaction-level labelling, \citet{Safar2026} provide a pipeline to allow for leakage-resistant\footnote{Following \citet{Safar2026}, we call a setting leakage-resistant when it closes the known leakage channels, here the split and the feature computation, without claiming that no leakage remains. We use this term throughout, including for similar settings that other authors name differently.} transaction-level feature construction. For transaction-level model training, illicit addresses can appear in both the training and test periods. To close this leakage channel, \citet{Safar2026} suggest shifting the analysis unit from transaction to address, because each address then appears only once. Focusing on addresses is also aligned with how security research and law enforcement treat illicit activity on Bitcoin: \citet{Meiklejohn2013} trace illicit funds by clustering addresses into the entities that control them, ransom payments are traced and measured through the addresses of ransomware operators \citep{Huang2018, PaquetClouston2019}, labelled datasets for ransomware detection are built at the address level \citep{Akcora2020}, and sanctions designate specific Bitcoin addresses \citep{OFAC2018}.

\citet{ElmougyLiu2023} took the pioneering step when they released Elliptic++, the first address-level extension of Elliptic, on which a random forest (RF) reaches an F1 of 0.845 on the illicit class. However, the shifting of the analysis unit from transaction to address does not solve the leakage problem. In the Elliptic++ dataset, the released split is positional rather than temporal and its address-level features are lifetime summaries of each address’s full history. This paper addresses both. We (1) recompute the Elliptic++ address-level features in a leakage-resistant way and share the exact definitions of them, (2) quantify the negative effect of leakage by comparing classification performance under leaky and leakage-resistant settings, (3) release the recomputed features and the construction code\footnote{Anonymised link for review: \url{https://github.com/DukHaanYumCha/Leakage-Resistant-Elliptic_pp}} to encourage further re-evaluation of results built on Elliptic++ under leakage-resistant settings.

\section{The Leakage Problem}
\label{sec:leakage}

\subsection{Current studies and their findings}
\label{subsec:datasets}

Empirical studies of Bitcoin began with the public transaction graph itself. Early work parsed the full ledger to describe its network structure \citep{RonShamir2013, KondorEtAl2014} and clustered addresses into the entities that control them \citep{Meiklejohn2013}. As labelled data became available, the focus moved from describing the network to detecting illicit activity in it, a task now often described as supervised classification. Two benchmark datasets dominate this field, the Elliptic dataset \citep{WeberEtAl2019} and its address-level extension Elliptic++ \citep{ElmougyLiu2023}.

The original Elliptic dataset \citep{WeberEtAl2019} provides a set of transactions with licit or illicit labels and attaches features to each transaction. These include local features (e.g., number of inputs and outputs, transaction volume) and aggregated features that summarise the same local information over the one-hop neighbouring transactions (maximum, minimum, standard deviation, and correlation coefficients). However, due to intellectual property concerns, these features are anonymised as numerical indices. Elliptic++ \citep{ElmougyLiu2023} extends the original dataset in two directions. Firstly, at the transaction level, $17$ additional features are appended, derived from the addresses participating in each transaction. Furthermore, over $822{,}000$ wallet addresses are recovered from the original Elliptic transactions, each described by $55$ summary statistics of local transaction activity. These address-level features are released with verbal descriptions but without exact definitions, so their computation cannot be reproduced from the release alone. 

Methodological work in this literature focuses on two families, graph neural networks (GNNs), which learn node representations by aggregating information along graph edges, and tree-based ensembles. The original Elliptic study applied graph convolutional networks (GCN) \citep{KipfWelling2017} alongside random forests. Subsequent work introduced temporal adaptations (EvolveGCN) \citep{ParejaEtAl2020}, attention-based architectures (GAT) \citep{VelickovicEtAl2018}, models capable of generalising to unseen nodes (GraphSAGE) \citep{HamiltonYingLeskovec2017}, unsupervised structural discrimination \citep{LoEtAl2023}, heterogeneous graph neural networks with metapath-based aggregation \citep{SongGu2023}, subgraph representation learning \citep{BelleiEtAl2024}, and heterogeneous graph transformers \citep{PerezCanoJurado2025}. Reported accuracy across these models is generally high. However, a recurring result is that tree-based models remain hard to beat. For example, \citet{ShaizatMussiraliyeva2025} show that a random forest (RF) fitted on a subset of Elliptic and Elliptic++ features selected by a specialised search algorithm is competitive with more advanced graph models. In the Elliptic++ study, \citet{ElmougyLiu2023} show that RF outperforms logistic regression alongside other tree- and GNN-based models. Similarly, \citet{NguyenBoersmaAcar2025} show that both RF and extreme gradient boosting baselines slightly outperform their proposed SAGE-FIN model. RF is thus the strongest model reported on Elliptic++, and it is the model whose performance we re-evaluate under leakage-resistant settings in \Cref{sec:performance}.

Recent work has raised concerns about the evaluation settings behind the high accuracy reported for Bitcoin classifiers \citep{Maganti2026, KhaleghpourMcKinney2026, Safar2026}. Both leakage channels introduced in \Cref{sec:intro} are named in this literature. Through the split channel, much of the literature evaluates on random or quasi-random splits, in which samples from the same time periods fall on both sides of the train/test split \citep{ElmougyLiu2023, NguyenBoersmaAcar2025}. Samples drawn from the same blockchain regime share similar feature characteristics, so a classifier can inflate its performance by recognising the current regime rather than by detecting illicit activity in a later one. Through the feature channel, static features computed over the full sampling period contain post-cutoff information by construction, and this leakage persists even under a temporal split. Both channels cast doubt on the reported performance under out-of-sample detection.

The transaction-level Elliptic literature has begun to treat this as a major concern. \citet{Maganti2026} compares evaluation settings directly across matched runs, holding other configurations (architecture, optimiser, loss, and seed) fixed. The study surprisingly finds that GraphSAGE \citep{HamiltonYingLeskovec2017} underperforms severely under a leaky setting compared to a leakage-resistant one\footnote{\citet{Maganti2026} calls this setting leakage-free.}. Leaking the graph structure in the test period to the model during training actively degrades performance. \citet{Maganti2026} also shows that an RF on raw features beats every graph model tested. In contrast, other recent models evaluated on Elliptic's temporal split \citep{AsiriSomasundaram2025, LinEtAl2026, DuyEtAl2026} do not isolate how much of their reported performance depends on the evaluation setting. \citet{KhaleghpourMcKinney2026} approach the problem from the feature side, proposing a general guideline for leakage-resistant feature computation\footnote{\citet{KhaleghpourMcKinney2026} call their guideline and features leakage-safe.} for temporal transaction networks and demonstrating it on the original transaction-level Elliptic dataset. They construct graphs and compute features using only edges observed up to each time step. Under a strict temporal split, an RF classifier reaches a ROC-AUC of 0.85 and an average precision of 0.54, but this performance comes almost entirely from Elliptic's original transaction features. The leakage-resistant graph features alone reach an average precision of only 0.05, and adding them to the original features leaves it unchanged. Working earlier in the same direction, \citet{NichollsEtAl2023FraudLens} similarly found that an RF trained on raw features outperforms several GNN baselines once the data split respects chronological order. Concurrently, \citet{Safar2026} audit the construction of Elliptic itself. They reverse-engineer most of the features, and show that some features are computed once with information that includes the test period, that these features link training and test transactions through shared addresses, and that the normalisation procedure used statistics of the whole dataset. They also note that the address-level information added by Elliptic++ was already largely present among Elliptic’s local features. This overlap was difficult to see at the time, since Elliptic’s features are released without clear definition or computation process. As a solution to improve Elliptic, they propose leakage-resistant features and splits that share no addresses, and release a tool for constructing such datasets. Their finding also limits what some transaction-level re-evaluations can show. \citet{Maganti2026} and \citet{KhaleghpourMcKinney2026} both still use Elliptic's original features, which contain the leaky address aggregates, so neither a strict inductive protocol nor leakage-resistant graph features layered on top of them makes the whole setting leakage-resistant by themselves. Together, these studies advocate leakage-resistant evaluation design and show that results reported on the Elliptic family need to be re-evaluated under it.

Same as for studies utilising the Elliptic dataset, also for Elliptic++ based studies, frequently a positional split is utilised, leading to high performance metrics \citep{NguyenBoersmaAcar2025, Song2025, ElmougyLiu2023}.  
However, only a temporal split would allow for a leakage-resistant evaluation. The strong depreciation in performance when switching from a positional to temporal split reported for the Elliptic dataset, calls for a re-evaluation of performance metrics achieved under positional split in the Bitcoin Forensics literature. 

In summary, leakage-resistant design has been advocated for the transaction-level data \citep{KhaleghpourMcKinney2026, Safar2026}, whereas the address level, the natural next step given the leakage channel of repeated addresses identified by \citet{Safar2026}, has seen neither a leakage-resistant reconstruction of its features nor a re-evaluation of its reported performance. This paper provides both for the address-level Elliptic++.

\subsection{Public ledger data}
\label{subsec:ledger-data}

This paper takes Elliptic++ as the starting point, which supplies the list of labelled addresses, the licit and illicit labels, the Elliptic time step of each address, and the values of 55 summary features of local transaction activity.

An address-level evaluation is leakage-resistant when both leakage channels of \Cref{sec:intro} are closed. Closing the split channel requires a temporal split, with a cutoff placed on the time line of the blockchain so that every training address is observed before the test period starts. Closing the feature channel requires that each address's features be computed only from its transactions up to its observation point, which in turn requires the individual transactions of the address, their amounts, and the blocks in which they were confirmed. The second requirement also presupposes exact feature definitions, since a feature can only be recomputed up to a cutoff if its definition is known. Once these conditions hold, the leakage channel of repeated addresses, identified by \citet{Safar2026}, is closed as well. Each address enters the sample once, on one side of the split, and its features summarise its own activity rather than aggregates shared with other samples.

Elliptic++ does not meet these requirements on its own. The time steps in its wallet file suffice to place each address on one side of a temporal split, but the released split is positional, and two properties of the Elliptic family prevent the features from being recomputed up to a cutoff. First, the underlying transactions are anonymised. The original Elliptic dataset releases its transactions as numerical indices rather than on-chain identifiers (i.e. transaction hashes), and Elliptic++ inherits this, releasing the 55 feature values but not the per-transaction amounts and block heights from which they are computed. A leakage-resistant recomputation, which restricts each address to the transactions observed at or before a temporal cutoff, requires those raw transactions together with their block heights.

Second, the features are described but not defined. Elliptic++ gives a verbal description of each of the 55 features, yet it does not provide their exact mathematical form or the per-transaction conventions needed to reproduce them on-chain. The verbal descriptions in what follows are taken from \citet{ElmougyLiu2023}, but the precise mathematical definitions, including the per-transaction conventions and the handling of edge cases such as self-transfers and fee attribution, are formalised in this paper and listed in full in \Cref{app:ael55-defs}.

Two further public datasets supply what is missing, a full Bitcoin ledger and a deanonymisation map. The ledger data come from the Bitcoin transaction network deposited by \citet{KondorDataset2021}, a full parse of the Bitcoin blockchain extracted with a modified Bitcoin Core client and originally built to study the structure of the Bitcoin transaction network \citep{KondorEtAl2014}. The deposit holds several releases, and we use the 2020 release, which covers every transaction up to February 2020 and therefore the whole Elliptic window from January 2016 to October 2017. Blocks, transactions, and addresses carry integer identifiers, and the release is organised as relational tables: a block table with timestamps, a transaction table that places each transaction in its block, input and output tables that record the address and the amount of every transaction input and output, and dictionaries that map the integer identifiers back to on-chain transaction hashes and address strings. The release lists $609{,}963{,}452$ addresses and about $1.22$ and $1.33$ billion transaction inputs and outputs. These tables provide exactly what the Elliptic family withholds. The amount each address sends to or receives from a transaction determines the value features, and the block of each transaction determines the timing features and places the transaction before or after the cutoff, so that each address's history can be truncated at its observation point, which closes the feature channel. 

To recompute the Elliptic++ features in a leakage-resistant manner, we have to link the anonymised Elliptic transactions with the Kondor ledger data. We follow the procedure of \citet{ElmougyLiu2023} and use the public deanonymisation map released on Kaggle \citep{AlexbenzikKaggle}. This map links each anonymised Elliptic transaction identifier to its on-chain transaction hash. Its author located the Elliptic features for transaction volume, fee, and the numbers of inputs, outputs, and unique addresses, and matched them against the blockchain using the known collection window and graph structure of the dataset \citep{Safar2026}. The map covers $202{,}804$ of the $203{,}769$ Elliptic transactions ($99.53\%$), and every mapped transaction is found in the Kondor ledger. Because Elliptic++ reuses Elliptic's transaction identifiers, the map joins the Elliptic++ addresses and time steps to Kondor's transactions and blocks. It restricts the ledger to the Elliptic transactions, so that features are recomputed over the same transaction set as the original construction, and it assigns every Elliptic time step its range of blocks, from which the cutoff block of the temporal split is read. The map closes no leakage channel on its own, but without it neither the feature channel nor the split channel could be closed at the address level.

\section{Leakage-Resistant Construction}
\label{sec:datasetnew}

\begin{figure*}[t]
  \centering
  \includegraphics[width=\textwidth]{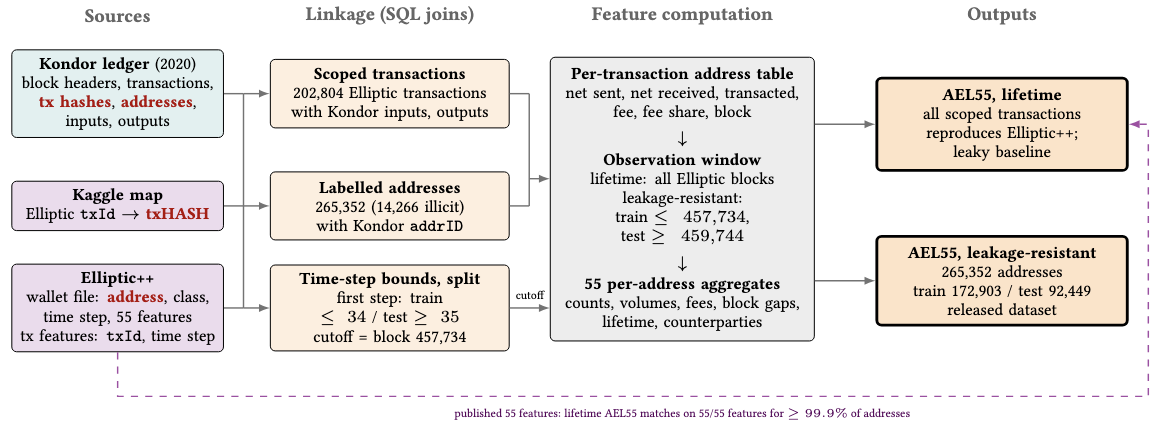}
  \caption{Dataset construction. Teal marks the Kondor ledger, violet the Elliptic family and the Kaggle map, and orange the produced data; red marks the cross-dataset join keys, and the dashed arrow carries the released Elliptic++ features used for validation. AEL55 refers to the recomputed 55 features from the Elliptic++ dataset.}
  \Description{Left-to-right flow chart in four columns. Sources: the Kondor ledger tables, the Kaggle deanonymisation map, and the Elliptic++ wallet and transaction files. Linkage: SQL joins on transaction hash and address produce 202,804 Elliptic transactions, 265,352 labelled addresses, and the time-step bounds with the temporal split and cutoff block 457,734. Feature computation: a per-transaction address table is restricted to an observation window, either lifetime or leakage-resistant, and aggregated into 55 per-address features. Outputs: a lifetime feature set that reproduces Elliptic++ and is validated against its released features, and a leakage-resistant feature set forming the released dataset with 172,903 training and 92,449 test addresses.}
  \label{fig:lineage}
\end{figure*}

This section describes how the Elliptic++ files, the deanonymisation map, and the Kondor ledger are joined into a single address-level dataset with a temporal split (\Cref{subsec:linkage}), and how the 55 features are recomputed in a leakage-resistant way (\Cref{subsec:features}). The validation of the recomputed features against the released values is shown later in \Cref{subsec:reproduction}. \Cref{fig:lineage} traces the full pipeline. Every step is implemented as SQL in DuckDB \citep{Raasveldt2019DuckDB}, which queries Parquet files directly without loading them into memory, with PyArrow and pandas used for file conversion and bookkeeping. The pipeline was executed on a Snellius\footnote{The Dutch national supercomputer, https://www.surf.nl/diensten/rekenen/snellius-de-nationale-supercomputer} compute node with 192 threads and 336\,GB of memory.

\subsection{Linking Elliptic++ to the ledger}
\label{subsec:linkage}

From the Kondor 2020 release \citep{KondorDataset2021}, six compressed files are used, namely the address dictionary, the transaction-hash dictionary, the block table, the transaction table, and the input and output tables. The block table records the height, hash, timestamp, and number of transactions of each block. The transaction table records the identifier of each transaction, its block, and its numbers of inputs and outputs. The output table records, for each transaction output, its transaction, its position, the receiving address, and the amount in satoshis, and the input table records the same for each input together with the earlier output it spends. The two largest tables, the inputs and the outputs, hold $1.22$ and $1.33$ billion rows, and the address dictionary holds $609{,}963{,}452$ addresses.

The labels enter through the addresses. The $265{,}354$ labelled addresses of Elliptic++, of which $14{,}266$ are illicit and $251{,}088$ licit, are matched to the Kondor address dictionary by an exact left join on the address string. The join matches all $14{,}266$ illicit and $251{,}086$ licit addresses, resulting in a class imbalance of $5.4\%$ illicit share. The remaining two licit addresses cannot be matched because the Elliptic++ release lists them as truncated strings rather than complete Bitcoin addresses, and they are dropped. Each matched address thereby receives its Kondor address identifier, through which all of its inputs and outputs in the ledger can be retrieved.

The transactions enter through the deanonymisation map \citep{AlexbenzikKaggle}, which provides transaction hashes for $202{,}804$ of the $203{,}769$ Elliptic transactions. The remaining $965$ have no entry in the map and therefore, following the approach of \citet{ElmougyLiu2023}, they are not covered in the analysis. The $202{,}804$ pairs of Elliptic transaction identifier and transaction hash in the map are joined in three steps: from the hash to the Kondor transaction identifier through the hash dictionary, from the transaction identifier to its block through the transaction table, and from the block to its timestamp through the block table. All $202{,}804$ transactions are found in the ledger. Every later feature computation is restricted to these transactions, matching the transaction set over which the Elliptic++ features were originally computed. 

The same procedure identifies the block range for each Elliptic time step. Elliptic++ aggregates transactions to one of 49 time-dependent samples, and joining these assignments to the Kondor dataset through the deanonymisation map gives, for each time sample, its first and last block, and their timestamps. \Cref{tab:elliptic-scope} reports the resulting training and test periods. The Elliptic window runs from block $391{,}200$ on 1 January 2016 to block $487{,}975$ on 2 October 2017, and the cutoff is set at the last block of time step 34, block $457{,}734$, which ends the training period on 18 March 2017.

\begin{table}[b]
\caption{Training and test periods on the Bitcoin block axis.}
\label{tab:elliptic-scope}
\resizebox{\columnwidth}{!}{%
\begin{tabular}{l cc rrr}
\toprule
Period & Blocks & Dates & Bitcoin txs & Elliptic txs & Elliptic addr. (illicit) \\
\midrule
Training & 391{,}200--457{,}734 & 2016-01-01 -- 2017-03-18 & 104{,}577{,}292 & 135{,}864 & 172{,}903 (9{,}377) \\
Test     & 459{,}744--487{,}975 & 2017-03-31 -- 2017-10-02 &  49{,}383{,}660 &  66{,}940 &  92{,}449 (4{,}889) \\
\midrule
Window   & 391{,}200--487{,}975 & 2016-01-01 -- 2017-10-02 & 157{,}574{,}585 & 202{,}804 & 265{,}352 (14{,}266) \\
\bottomrule
\end{tabular}}

\smallskip
\begin{minipage}{\linewidth}\footnotesize
\textit{Note:} Training covers Elliptic time steps 1 to 34 and testing time steps 35 to 49; the cutoff is the last training block, 457{,}734. Bitcoin transactions are counted from the Kondor block table and include coinbase transactions; the window also includes the 2{,}009 blocks between the two periods, which contain no Elliptic transaction. Elliptic transactions are those recovered by the deanonymisation map. Addresses are assigned to a period by their first appearance.
\end{minipage}
\end{table}

\subsection{Feature recomputation}
\label{subsec:features}

The 55 features are recomputed from the Elliptic transactions, following the verbal descriptions of \citet{ElmougyLiu2023}. As in the original construction, they summarise each address's activity within the Elliptic transactions rather than its full history on the chain. The computation has two stages, a per-transaction table that records the stake of each labelled address in every transaction it takes part in, with 367{,}517 rows, and an aggregation of that table to one row per address. 

The per-transaction table records, for each labelled address and each transaction it takes part in, the amount the address contributes as inputs and the amount it collects as outputs. An address that spends coins and receives change in the same transaction appears on both sides, so its stake is netted. The net amount sent is the excess of its inputs over its outputs. The net amount received is the excess of its outputs over its inputs, and the amount transacted is the absolute difference between the two. A transaction counts as a sending transaction of the address if the net amount sent is positive, and as a receiving transaction if the net amount received is positive. The fee of the transaction is the excess of its total input over its total output, and the address is attributed a share of it in proportion to the larger of its input and output amounts, relative to the total input of the transaction. All amounts are in bitcoin, and every row carries the block of its transaction.

Each of these five per-transaction quantities, the amounts transacted, sent, and received, the fee, and the fee share, is reduced over the transactions of an address to five summary statistics: the total, minimum, maximum, mean, and median, which gives 25 value features. The remaining 30 features describe activity, timing, and counterparties. The three transaction counts record all transactions of the address and its sending and receiving transactions. The block gaps are the differences between the blocks of consecutive transactions of the address, taken over all transactions and over sending and receiving transactions separately, and are zero when fewer than two such transactions exist. The lifetime features record the first and last block of the address, their difference, the first sending and first receiving block, both zero when the address never sends or never receives, and the number of distinct time steps in which it is active. The counterparties of an address in a transaction are the addresses on the opposite side, the output addresses when it sends and the input addresses when it receives, excluding the address itself. The five summary statistics of the number of transactions shared with each counterparty, and the number of counterparties shared in more than one transaction, complete the 55 features. \Cref{tab:ael55-defs} in \Cref{app:ael55-defs} gives the formula of each feature.

The two output feature sets in \Cref{fig:lineage} come from the same computation and differ only in which rows of the per-transaction table enter the aggregation. The lifetime set uses every Elliptic transaction of each address, which reproduces the construction of Elliptic++ and serves as the leaky baseline. The leakage-resistant set keeps, for each training address, only its transactions up to the cutoff block, and for each test address its transactions up to the end of the test period. Every feature is therefore computable at the observation point of its address, and since models are fitted or scored immediately after these points, the legitimacy rule of \citet{Kaufman2012} holds. 

Some features are identical or linearly dependent by construction. The total block gap equals the lifetime, since the gaps between consecutive transactions of an address sum to the difference between its last and first block, and the lifetime is itself the difference between the last and first block. The totals and means of the amounts transacted, sent, and received satisfy \textit{transacted $=$ sent $+$ received} for every address, and the number of transactions equals the numbers of sending and receiving transactions combined. \Cref{tab:identities} lists these identities and the share of training addresses that satisfy them in the features released by \citet{ElmougyLiu2023} and the leakage-resistant features computed here.

\begin{table}[t]
\caption{Exact identities among the features and the share of training addresses that satisfy them.}
\label{tab:identities}
\small
\begin{tabular}{@{}l rr@{}}
\toprule
Identity & Released & Leakage-resistant \\
\midrule
Total block gap $=$ lifetime                     & 99.971\% (50) & 100\% \\
Lifetime $=$ last block $-$ first block          & 100\%         & 100\% \\
Transacted $=$ sent $+$ received, totals         & 99.998\% (4)  & 100\% \\
Transacted $=$ sent $+$ received, means          & 99.997\% (5)  & 100\% \\
Transactions $=$ sending $+$ receiving           & 100\%         & 99.998\% (3) \\
\bottomrule
\end{tabular}

\smallskip
\begin{minipage}{\linewidth}\footnotesize
\textit{Note:} Shares of the 172{,}903 training addresses whose features satisfy each identity, with the number of exceptions in parentheses. Both sides of an identity, say $a$ and $b$, are computed from the features of the same address. Integer-valued features satisfy an identity only if the two sides are equal, $\lvert a - b \rvert = 0$. Amounts in bitcoin satisfy it if $\lvert a - b \rvert \le 10^{-8} + 10^{-6}\lvert b \rvert$, a relative tolerance of $10^{-6}$ with an absolute floor of one satoshi ($10^{-8}$ bitcoin).
\end{minipage}
\end{table}

The exceptions have two sources. In the released features, the total block gap exceeds the lifetime for 50 training addresses, while their lifetimes agree with ours. Gaps between consecutive transactions can sum to more than the difference between the last and first block only if the transactions are not ordered by block when the gaps are taken, so these values are consistent with an ordering error in the original computation of \citet{ElmougyLiu2023}. The released amounts also violate \textit{transacted $=$ sent $+$ received} for 4 addresses in the totals and 5 in the means, which our recomputation does not reproduce. In the leakage-resistant features, three training addresses each have a transaction in which they contribute exactly what they collect. Its net stake is zero, so it counts towards the total but as neither a sending nor a receiving transaction. \citet{ElmougyLiu2023} appear to count such a transaction as sending, since for the five addresses with such transactions in the lifetime set the released sending count exceeds ours by one or two.

\begin{figure*}[t]
  \centering
  \includegraphics[width=\textwidth]{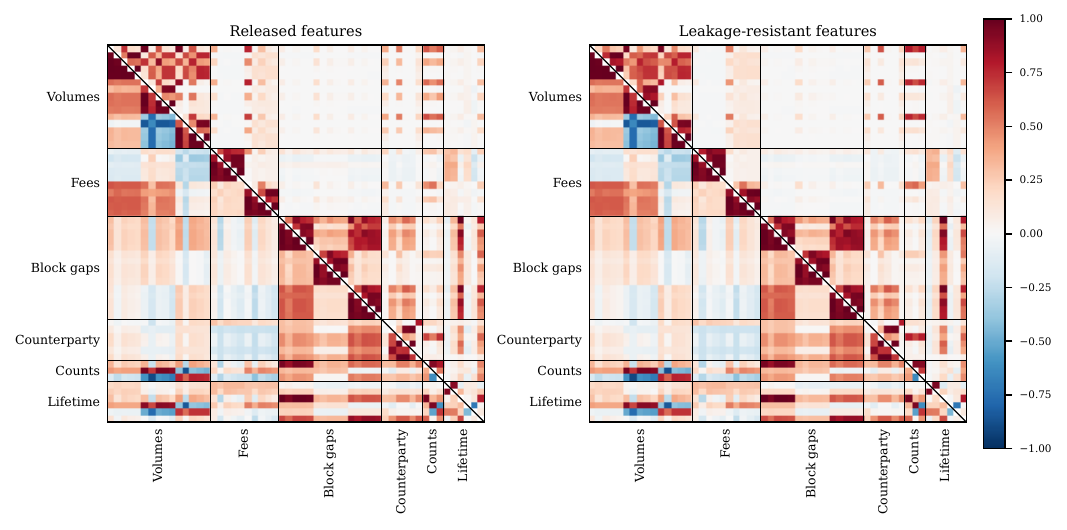}
  \caption{Correlations between the 55 features on training addresses, for the released features (left) and the leakage-resistant features (right). In each panel, Pearson correlations are shown above the diagonal and Spearman correlations below it. Features follow the order of \Cref{tab:ael55-defs} in \Cref{app:ael55-defs} of the Appendix, and lines separate the feature families.}
  \Description{Two square heatmaps side by side, each with 55 rows and columns grouped into six families: volumes, fees, block gaps, counterparty, counts, and lifetime. Colours run from blue for negative to red for positive correlation. Each panel shows one feature set correlated with itself, Pearson above the diagonal and Spearman below. Both panels show strong positive blocks within the volume, fee, and block-gap families and are nearly identical.}
  \label{fig:corr55}
\end{figure*}

\Cref{fig:corr55} shows the correlations among the 55 features on the training addresses, for the released features in the left panel and the leakage-resistant features in the right. Each panel correlates one feature set with itself, with Pearson correlations above the diagonal and Spearman correlations below it. The dark blocks along the diagonal reflect the redundancy of the summary statistics, the total, minimum, maximum, mean, and median of the same quantity. Spearman correlations among the five statistics of a quantity average 0.85 to 0.99 for the volumes, fees, and block gaps, since most addresses have few transactions and their totals, extremes, means, and medians often coincide. The amounts sent and received are negatively rank-correlated (Spearman $-0.35$), since many addresses only send or only receive within the Elliptic transactions, yet positively correlated in value (Pearson $0.88$), since a few high-volume addresses both receive and send large amounts. Given the high correlation and even exact duplication among many of these $55$ features, including them all in the feature set results in multicollinearity, which one should take into account. 

Comparing the two panels shows that the cutoff hardly changes the correlation structure, both the redundancy within each family and the relations between families. No Spearman correlation changes by more than 0.05, and Pearson correlations, which weigh extreme values more heavily, change by up to 0.38, reflecting the 2{,}501 training addresses whose features the cutoff alters. The two panels, however, differ in two respects, the recomputation and the cutoff. The lifetime set separates the two, since it is recomputed in the same way but without the cutoff, so its correlations differ from those of the released features through the recomputation alone. They agree to within 0.006 for Spearman and 0.013 for Pearson. The recomputation therefore barely changes the correlations, and the differences between the two panels can be attributed to the cutoff.

\section{Model Performance}
\label{sec:performance}

\subsection{Evaluation setup}
\label{subsec:setup}

The evaluation follows the classification pipeline of \citet{ElmougyLiu2023}, so that differences in performance can be attributed to the data rather than to the models. Two classifiers are fitted, a logistic regression (LR) with the default ridge penalty and a random forest (RF) with 50 trees, both without class weighting. Features are MinMax-scaled, and predictions use the default decision threshold of 0.5. Performance is reported as precision, recall, and F1 on the illicit class. LR is fitted once with the default L-BFGS solver and L2 penalty of scikit-learn, following \citet{ElmougyLiu2023}. They fix no random seed for the RF, so its performance is reported as the mean over 10 seeds, with the standard deviation reported below. Under the positional split, the scaler is fitted on all labelled addresses. Under the temporal split, it is fitted on training addresses only, so that no statistic of the test period enters training.

\subsection{Reproduction}
\label{subsec:reproduction}

We first validate the recovered lifetime features against the released values for every labelled address. Each of the 55 features matches the released value to high precision\footnote{A recomputed value $c$ matches a released value $p$ if $\lvert c - p \rvert \le 10^{-8} + 10^{-6}\lvert p \rvert$, a relative tolerance of $10^{-6}$ with an absolute floor of $10^{-8}$, which is one satoshi for amounts in bitcoin. For integer-valued features, the bound is below one even for the largest value in the data, $487{,}975$, so the condition requires exact equality, $\lvert c - p \rvert = 0$.} for at least 99.9\% of the 265{,}352 addresses, and the least-matching feature, the total counterparty count, still matches for 99.915\%. The fee share is the one feature whose definition departs from its verbal description.\footnote{\citet{ElmougyLiu2023} describe the fee share as ``total fees as share of BTC transacted'', which reads as a single ratio of an address's total fees to its total amount transacted. The released values instead follow a per-transaction definition. In each transaction, the address is attributed the part of the fee proportional to the larger of its input and output amounts, relative to the total input of the transaction, and these per-transaction shares are then summarised over the address's transactions with the same five statistics as the other value features (\Cref{tab:ael55-defs} in \Cref{app:ael55-defs}).} Computed per transaction and then aggregated, it matches the released values within one percent\footnote{Within one percent means $\lvert c - p \rvert \le 10^{-4} + 0.01\lvert p \rvert$, with amounts in bitcoin. \Cref{app:validation} in the Appendix justifies the absolute term. Replacing it with one satoshi flags 51 further addresses, all differing by less than 0.0003 bitcoin, and excluding them leaves the RF and LR results unchanged.} for 99.99\% of addresses, against 7.65\% when computed as a single ratio of the address's total fees to its total amount transacted. A further 86 addresses (0.03\%) differ from the released values by more than one percent on at least one of the 25 value features, but dropping them leaves model performance unchanged (\Cref{tab:divergent-sensitivity} in \Cref{app:validation}).

\begin{table}[t]
\caption{Reproduction of the Elmougy and Liu wallet classifier on the illicit class.}
\label{tab:reproduction}
\resizebox{\columnwidth}{!}{%
\begin{tabular}{l ccc ccc ccc}
\toprule
& \multicolumn{3}{c}{Recovered} & \multicolumn{3}{c}{Released} & \multicolumn{3}{c}{\citet{ElmougyLiu2023}} \\
\cmidrule(lr){2-4}\cmidrule(lr){5-7}\cmidrule(lr){8-10}
Model & P & R & F1 & P & R & F1 & P & R & F1 \\
\midrule
LR & 0.492 & 0.062 & 0.110 & 0.499 & 0.062 & 0.110 & 0.477 & 0.046 & 0.083 \\
RF & \sd{0.911}{0.004} & \sd{0.786}{0.003} & \sd{0.844}{0.002}
   & \sd{0.916}{0.004} & \sd{0.787}{0.003} & \sd{0.846}{0.002}
   & 0.911 & 0.789 & 0.845 \\
\bottomrule
\end{tabular}}

\smallskip
\begin{minipage}{\linewidth}\footnotesize
\textit{Note:} LR, logistic regression; RF, random forest with 50 trees. Recovered features are recomputed from the Kondor ledger as lifetime summaries; released features are those of \citet{ElmougyLiu2023}. Both follow their pipeline, with MinMax scaling over the 265{,}352 labelled addresses and the original positional 70/30 split. \citet{ElmougyLiu2023} fix no random seed for RF, and their figures are from a single run; RF here is reported as the mean over 10 seeds, with the standard deviation below. Precision, recall, and F1 are at the default 0.5 threshold.
\end{minipage}
\end{table}

\Cref{tab:reproduction} reports the Elliptic++ wallet classifier under its original positional 70/30 split. On the recovered lifetime features, the RF reaches an F1 of 0.844, against 0.846 on the released features and 0.845 reported by \citet{ElmougyLiu2023}, with a standard deviation of 0.002 across seeds. LR reaches 0.110 on both recovered and released feature sets, above the 0.083 reported in the original study. We compute the features for the labelled addresses only and therefore fit the MinMax scaler on them, whereas \citet{ElmougyLiu2023} fit it on all wallet addresses, including those of unknown class. The choice does not affect RF, whose splits are unchanged by a monotone rescaling of each feature, and barely affects LR. On the released features, scaling over all addresses including the unknown ones gives an F1 of 0.843 for RF and 0.102 for LR, against 0.846 and 0.110 when scaling over the labelled addresses. Dropping the 86 diverging addresses leaves both models unchanged (\Cref{tab:divergent-sensitivity} in \Cref{app:validation}). The recovered features therefore reproduce the released ones in their values, their correlations, and the model results they yield, and serve as the baseline for the comparisons that follow.

\subsection{Effect of leakage}
\label{subsec:leakage-effect}

\Cref{tab:leakage} separates the two leakage channels. The first column group repeats the released-feature results of \Cref{tab:reproduction}. The second keeps the released, leaky features but replaces the positional split with the temporal split of \Cref{subsec:linkage}, which closes the split channel alone. The third also replaces the features with the leakage-resistant ones, which closes both channels.

Closing the split channel leads to notable changes in terms of Precision, Recall, and F1 score. The RF F1 falls from 0.846 to 0.257, as precision drops from 0.916 to 0.381 and recall from 0.787 to 0.196. The drop is not an effect of a change in the illicit share, which is almost the same in the two periods, 5.4\% in training and 5.3\% in testing (\Cref{tab:elliptic-scope}). Under the positional split, training and test addresses come from the same periods and share their activity patterns, so the RF can score well by matching addresses to patterns it has already seen in that period. Once the test addresses all appear after the training period, this advantage disappears. The RF also becomes more unstable, with the standard deviation of its F1 across seeds rising from 0.002 to 0.035. LR moves in the opposite direction, from 0.110 to 0.288, with lower precision but much higher recall, and now matches or exceeds the RF.

Utilising the leakage-resistant features, hence closing the feature channel, results in changes in the features of 1.45\% of training addresses, and of 0.52\% of the illicit ones. In Elliptic++, most addresses are active for too short a time to accumulate post-cutoff information in their lifetime summaries. Consequently, we do not expect drastic changes in the Precision, Recall, and F1 score, when utilising the leakage-resistant features. Indeed, LR's performance scores hover around the values observed under the temporal split with the released features. However, for the RF we observe an improvement in Precision, Recall, and F1 score when utilising the leakage-resistant features. Notably, the standard deviation also increases; however, the average score improves across all three metrics. A much stronger improvement would have been surprising, given that the features of only a limited number of addresses were exposed to future information leakage. 

This contrasts with Elliptic transaction-level studies, whose features of address aggregates summarise each address's activity on the whole chain up to a block well beyond the test period, so that almost every transaction carries look-ahead information \citep{Safar2026} and the original features unsurprisingly dominate leakage-resistant features added to the feature set \citep{KhaleghpourMcKinney2026}. 

However, the usability of the leakage-resistant features developed in this paper naturally extends beyond Elliptic++. With the equations for their computation and the leakage-resistant pipeline provided, leakage-resistant features for any set of labelled Bitcoin addresses can be obtained. We showe that even for a small percentage of addresses affected by future information leakage, a notable effect on performance scores is visible. Furthermore, we showed the drastic effects of not using a temporal split during model training and evaluation. Under a temporal evaluation, the RF F1 is about 0.26 to 0.29 rather than 0.85 under the positional split, and the advantage of the RF over LR disappears. 

\begin{table}[t]
\caption{Effect of information leakage on performance.}
\label{tab:leakage}
\resizebox{\columnwidth}{!}{%
\begin{tabular}{l ccc ccc ccc}
\toprule
& \multicolumn{3}{c}{Positional split} & \multicolumn{3}{c}{Temporal split} & \multicolumn{3}{c}{Temporal split} \\
& \multicolumn{3}{c}{released features} & \multicolumn{3}{c}{released features} & \multicolumn{3}{c}{leakage-resistant features} \\
\cmidrule(lr){2-4}\cmidrule(lr){5-7}\cmidrule(lr){8-10}
Model & P & R & F1 & P & R & F1 & P & R & F1 \\
\midrule
LR & 0.499 & 0.062 & 0.110 & 0.199 & 0.520 & 0.288 & 0.198 & 0.506 & 0.285 \\
RF & \sd{0.916}{0.004} & \sd{0.787}{0.003} & \sd{0.846}{0.002}
   & \sd{0.381}{0.048} & \sd{0.196}{0.033} & \sd{0.257}{0.035}
   & \sd{0.407}{0.074} & \sd{0.228}{0.050} & \sd{0.291}{0.058} \\
\bottomrule
\end{tabular}}

\smallskip
\begin{minipage}{\linewidth}\footnotesize
\textit{Note:} LR, logistic regression; RF, random forest with 50 trees, mean over 10 seeds with the standard deviation below. The first group repeats the released column of \Cref{tab:reproduction}: positional 70/30 split, MinMax scaling over all labelled addresses. The temporal split trains on addresses first seen in Elliptic time steps 1--34 and tests on those first seen in steps 35--49, with MinMax scaling fitted on training addresses only. Leakage-resistant features cap each training address at the cutoff block. Precision, recall, and F1 at the default 0.5 threshold.
\end{minipage}
\end{table}

\section{Conclusion and Discussion}
\label{sec:conclusion}

This paper asks whether the strong performance reported for address-level classification on Elliptic++ reflects out-of-sample detection or information leakage, and the answer is largely leakage. We recover the exact definitions of the 55 Elliptic++ address-level features by linking Elliptic transactions to a full Bitcoin ledger through a public deanonymisation map. The recovered features reproduce the released values for at least 99.9\% of the labelled addresses and the reported RF performance to within 0.002 in F1. With these definitions, we recompute the features in a leakage-resistant way with a temporal split. Under this evaluation, the RF F1 on the illicit class falls from 0.846 to 0.291, and the RF no longer outperforms logistic regression. 

Two implications follow. Results and model rankings reported on Elliptic++ under its released split should be re-evaluated under a temporal split before they are read as evidence of detection ability. The leakage-resistant features close both known leakage channels, so they allow for model training and development under realistic settings. Furthermore, the released definitions and code allow the same features to be computed for addresses outside Elliptic++.

Two limitations remain, both following from our decision to keep the pipeline of \citet{ElmougyLiu2023} so that our results are directly comparable with theirs. First, all 55 features enter the classifiers, although many of them are highly correlated or exactly collinear (\Cref{fig:corr55}, \Cref{tab:identities}). The five summary statistics of a quantity often coincide for addresses with few transactions, and some features are identical by construction. Such redundancy leaves RF largely unaffected but makes the coefficients of LR unstable, so feature selection or dimension reduction should precede modelling in future use of the dataset. Second, LR and RF are fitted with default hyperparameters, without class weighting, and with the default decision threshold of 0.5. Baselines against which a proposed model is judged should be tuned with the same care as the model itself, so that any reported advantage reflects the model improvement.

\section*{Ethics and Privacy Statement}
This study uses only public data, the Bitcoin ledger, published as a parsed release under CC0 \citep{KondorDataset2021}, and the released Elliptic++ dataset \citep{ElmougyLiu2023} and deanonymisation map \citep{AlexbenzikKaggle}. It involves no human participants, and it attributes no address to a real-world identity beyond the pseudonymous addresses already released with Elliptic++. The main societal risk lies in the use of classification results. An address labelled illicit by a model carries statistical uncertainty. Due to false positives, model outputs should be scrutinised before supporting investigation and should not by themselves justify freezing funds or other action against an address's owner. By showing how leakage inflates detection performance, this paper aims to reduce the risk of deploying classifiers whose reliability has been overstated.

\newpage
\bibliographystyle{ACM-Reference-Format}
\bibliography{refs}


\begin{thebibliography}{38}


\ifx \showCODEN    \undefined \def \showCODEN     #1{\unskip}     \fi
\ifx \showISBNx    \undefined \def \showISBNx     #1{\unskip}     \fi
\ifx \showISBNxiii \undefined \def \showISBNxiii  #1{\unskip}     \fi
\ifx \showISSN     \undefined \def \showISSN      #1{\unskip}     \fi
\ifx \showLCCN     \undefined \def \showLCCN      #1{\unskip}     \fi
\ifx \shownote     \undefined \def \shownote      #1{#1}          \fi
\ifx \showarticletitle \undefined \def \showarticletitle #1{#1}   \fi
\ifx \showURL      \undefined \def \showURL       {\relax}        \fi
\providecommand\bibfield[2]{#2}
\providecommand\bibinfo[2]{#2}
\providecommand\natexlab[1]{#1}
\providecommand\showeprint[2][]{arXiv:#2}

\bibitem[Akcora et~al\mbox{.}(2020)]%
        {Akcora2020}
\bibfield{author}{\bibinfo{person}{Cuneyt~Gurcan Akcora}, \bibinfo{person}{Yitao Li}, \bibinfo{person}{Yulia~R. Gel}, {and} \bibinfo{person}{Murat Kantarcioglu}.} \bibinfo{year}{2020}\natexlab{}.
\newblock \showarticletitle{{BitcoinHeist}: Topological Data Analysis for Ransomware Prediction on the {Bitcoin} Blockchain}. In \bibinfo{booktitle}{\emph{Proceedings of the Twenty-Ninth International Joint Conference on Artificial Intelligence ({IJCAI})}}. \bibinfo{pages}{4439--4445}.
\newblock


\bibitem[{alexbenzik}(2019)]%
        {AlexbenzikKaggle}
\bibfield{author}{\bibinfo{person}{{alexbenzik}}.} \bibinfo{year}{2019}\natexlab{}.
\newblock \bibinfo{booktitle}{\emph{Deanonymized 99.5 pct of Elliptic Transactions}}.
\newblock
\urldef\tempurl%
\url{https://www.kaggle.com/datasets/alexbenzik/deanonymized-995-pct-of-elliptic-transactions}
\showURL{%
Retrieved October 13, 2026 from \tempurl}


\bibitem[Arp et~al\mbox{.}(2022)]%
        {Arp2022}
\bibfield{author}{\bibinfo{person}{Daniel Arp}, \bibinfo{person}{Erwin Quiring}, \bibinfo{person}{Feargus Pendlebury}, \bibinfo{person}{Alexander Warnecke}, \bibinfo{person}{Fabio Pierazzi}, \bibinfo{person}{Christian Wressnegger}, \bibinfo{person}{Lorenzo Cavallaro}, {and} \bibinfo{person}{Konrad Rieck}.} \bibinfo{year}{2022}\natexlab{}.
\newblock \showarticletitle{Dos and Don'ts of Machine Learning in Computer Security}. In \bibinfo{booktitle}{\emph{31st USENIX Security Symposium}} \emph{(\bibinfo{series}{USENIX Security '22})}. \bibinfo{pages}{3971--3988}.
\newblock


\bibitem[Asiri and Somasundaram(2025)]%
        {AsiriSomasundaram2025}
\bibfield{author}{\bibinfo{person}{Ahmad Asiri} {and} \bibinfo{person}{K. Somasundaram}.} \bibinfo{year}{2025}\natexlab{}.
\newblock \showarticletitle{Graph Convolution Network for Fraud Detection in {Bitcoin} Transactions}.
\newblock \bibinfo{journal}{\emph{Scientific Reports}}  \bibinfo{volume}{15} (\bibinfo{year}{2025}), \bibinfo{pages}{11076}.
\newblock
\href{https://doi.org/10.1038/s41598-025-95672-w}{doi:\nolinkurl{10.1038/s41598-025-95672-w}}


\bibitem[Bellei et~al\mbox{.}(2024)]%
        {BelleiEtAl2024}
\bibfield{author}{\bibinfo{person}{Claudio Bellei}, \bibinfo{person}{Muhua Xu}, \bibinfo{person}{Ross Phillips}, \bibinfo{person}{Tom Robinson}, \bibinfo{person}{Mark Weber}, \bibinfo{person}{Tim Kaler}, \bibinfo{person}{Charles~E. Leiserson}, \bibinfo{person}{Arvind}, {and} \bibinfo{person}{Jie Chen}.} \bibinfo{year}{2024}\natexlab{}.
\newblock \showarticletitle{The Shape of Money Laundering: Subgraph Representation Learning on the Blockchain with the {Elliptic2} Dataset}. In \bibinfo{booktitle}{\emph{KDD Workshop on Machine Learning in Finance (MLF '24)}}. \bibinfo{address}{Barcelona, Spain}.
\newblock


\bibitem[{Chainalysis}(2026)]%
        {chainalysis2026}
\bibfield{author}{\bibinfo{person}{{Chainalysis}}.} \bibinfo{year}{2026}\natexlab{}.
\newblock \bibinfo{title}{The 2026 Crypto Crime Report}.
\newblock
\shownote{\url{https://www.chainalysis.com/reports/crypto-crime-2026/}}.
\newblock


\bibitem[Christin(2013)]%
        {Christin2013}
\bibfield{author}{\bibinfo{person}{Nicolas Christin}.} \bibinfo{year}{2013}\natexlab{}.
\newblock \showarticletitle{Traveling the {Silk Road}: A Measurement Analysis of a Large Anonymous Online Marketplace}. In \bibinfo{booktitle}{\emph{Proceedings of the 22nd International Conference on World Wide Web}} \emph{(\bibinfo{series}{WWW '13})}. \bibinfo{publisher}{ACM}, \bibinfo{address}{Rio de Janeiro, Brazil}, \bibinfo{pages}{213--224}.
\newblock
\href{https://doi.org/10.1145/2488388.2488408}{doi:\nolinkurl{10.1145/2488388.2488408}}


\bibitem[Duy et~al\mbox{.}(2026)]%
        {DuyEtAl2026}
\bibfield{author}{\bibinfo{person}{Dang~Sy Duy}, \bibinfo{person}{Nguyen~Duy Chien}, \bibinfo{person}{Kapil Dev}, {and} \bibinfo{person}{Jeff Nijsse}.} \bibinfo{year}{2026}\natexlab{}.
\newblock \showarticletitle{Normalisation and Initialisation Strategies for Graph Neural Networks in Blockchain Anomaly Detection}.
\newblock \bibinfo{journal}{\emph{arXiv preprint arXiv:2602.23599}} (\bibinfo{year}{2026}).
\newblock


\bibitem[Elmougy and Liu(2023)]%
        {ElmougyLiu2023}
\bibfield{author}{\bibinfo{person}{Youssef Elmougy} {and} \bibinfo{person}{Ling Liu}.} \bibinfo{year}{2023}\natexlab{}.
\newblock \showarticletitle{Demystifying Fraudulent Transactions and Illicit Nodes in the {Bitcoin} Network for Financial Forensics}. In \bibinfo{booktitle}{\emph{Proceedings of the 29th ACM SIGKDD Conference on Knowledge Discovery and Data Mining}} \emph{(\bibinfo{series}{KDD '23})}. \bibinfo{publisher}{ACM}, \bibinfo{address}{Long Beach, CA, USA}, \bibinfo{pages}{3979--3990}.
\newblock
\href{https://doi.org/10.1145/3580305.3599803}{doi:\nolinkurl{10.1145/3580305.3599803}}


\bibitem[Foley et~al\mbox{.}(2019)]%
        {Foleyetal2019}
\bibfield{author}{\bibinfo{person}{Sean Foley}, \bibinfo{person}{Jonathan~R Karlsen}, {and} \bibinfo{person}{Tālis~J Putniņš}.} \bibinfo{year}{2019}\natexlab{}.
\newblock \showarticletitle{Sex, Drugs, and Bitcoin: How Much Illegal Activity Is Financed through Cryptocurrencies?}
\newblock \bibinfo{journal}{\emph{The Review of Financial Studies}} \bibinfo{volume}{32}, \bibinfo{number}{5} (\bibinfo{date}{05} \bibinfo{year}{2019}), \bibinfo{pages}{1798--1853}.
\newblock
\showISSN{0893-9454}
\showeprint{https://academic.oup.com/rfs/article-pdf/32/5/1798/28374143/hhz015.pdf}
\href{https://doi.org/10.1093/rfs/hhz015}{doi:\nolinkurl{10.1093/rfs/hhz015}}


\bibitem[Hamilton et~al\mbox{.}(2017)]%
        {HamiltonYingLeskovec2017}
\bibfield{author}{\bibinfo{person}{William~L. Hamilton}, \bibinfo{person}{Rex Ying}, {and} \bibinfo{person}{Jure Leskovec}.} \bibinfo{year}{2017}\natexlab{}.
\newblock \showarticletitle{Inductive Representation Learning on Large Graphs}. In \bibinfo{booktitle}{\emph{Advances in Neural Information Processing Systems}}, Vol.~\bibinfo{volume}{30}.
\newblock


\bibitem[Huang et~al\mbox{.}(2018)]%
        {Huang2018}
\bibfield{author}{\bibinfo{person}{Danny~Yuxing Huang}, \bibinfo{person}{Maxwell~Matthaios Aliapoulios}, \bibinfo{person}{Vector~Guo Li}, \bibinfo{person}{Luca Invernizzi}, \bibinfo{person}{Elie Bursztein}, \bibinfo{person}{Kylie McRoberts}, \bibinfo{person}{Jonathan Levin}, \bibinfo{person}{Kirill Levchenko}, \bibinfo{person}{Alex~C. Snoeren}, {and} \bibinfo{person}{Damon McCoy}.} \bibinfo{year}{2018}\natexlab{}.
\newblock \showarticletitle{Tracking Ransomware End-to-end}. In \bibinfo{booktitle}{\emph{2018 IEEE Symposium on Security and Privacy}} \emph{(\bibinfo{series}{SP '18})}. \bibinfo{publisher}{IEEE}, \bibinfo{address}{San Francisco, CA, USA}, \bibinfo{pages}{618--631}.
\newblock
\href{https://doi.org/10.1109/SP.2018.00047}{doi:\nolinkurl{10.1109/SP.2018.00047}}


\bibitem[Kapoor and Narayanan(2023)]%
        {Kapoor2023}
\bibfield{author}{\bibinfo{person}{Sayash Kapoor} {and} \bibinfo{person}{Arvind Narayanan}.} \bibinfo{year}{2023}\natexlab{}.
\newblock \showarticletitle{Leakage and the Reproducibility Crisis in Machine-Learning-Based Science}.
\newblock \bibinfo{journal}{\emph{Patterns}}  \bibinfo{volume}{4} (\bibinfo{year}{2023}), \bibinfo{pages}{100804}.
\newblock
\href{https://doi.org/10.1016/j.patter.2023.100804}{doi:\nolinkurl{10.1016/j.patter.2023.100804}}


\bibitem[Kaufman et~al\mbox{.}(2012)]%
        {Kaufman2012}
\bibfield{author}{\bibinfo{person}{Shachar Kaufman}, \bibinfo{person}{Saharon Rosset}, \bibinfo{person}{Claudia Perlich}, {and} \bibinfo{person}{Ori Stitelman}.} \bibinfo{year}{2012}\natexlab{}.
\newblock \showarticletitle{Leakage in Data Mining: Formulation, Detection, and Avoidance}.
\newblock \bibinfo{journal}{\emph{ACM Transactions on Knowledge Discovery from Data}} \bibinfo{volume}{6}, \bibinfo{number}{4}, Article \bibinfo{articleno}{15} (\bibinfo{year}{2012}).
\newblock
\href{https://doi.org/10.1145/2382577.2382579}{doi:\nolinkurl{10.1145/2382577.2382579}}


\bibitem[Khaleghpour and McKinney(2026)]%
        {KhaleghpourMcKinney2026}
\bibfield{author}{\bibinfo{person}{Hamideh Khaleghpour} {and} \bibinfo{person}{Brett McKinney}.} \bibinfo{year}{2026}\natexlab{}.
\newblock \showarticletitle{Leakage Safe Graph Features for Interpretable Fraud Detection in Temporal Transaction Networks}.
\newblock \bibinfo{journal}{\emph{arXiv preprint arXiv:2603.06632}} (\bibinfo{year}{2026}).
\newblock


\bibitem[Kipf and Welling(2017)]%
        {KipfWelling2017}
\bibfield{author}{\bibinfo{person}{Thomas~N. Kipf} {and} \bibinfo{person}{Max Welling}.} \bibinfo{year}{2017}\natexlab{}.
\newblock \showarticletitle{Semi-Supervised Classification with Graph Convolutional Networks}. In \bibinfo{booktitle}{\emph{International Conference on Learning Representations (ICLR)}}.
\newblock


\bibitem[Kondor et~al\mbox{.}(2014)]%
        {KondorEtAl2014}
\bibfield{author}{\bibinfo{person}{D{\'a}niel Kondor}, \bibinfo{person}{M{\'a}rton P{\'o}sfai}, \bibinfo{person}{Istv{\'a}n Csabai}, {and} \bibinfo{person}{G{\'a}bor Vattay}.} \bibinfo{year}{2014}\natexlab{}.
\newblock \showarticletitle{Do the Rich Get Richer? {An} Empirical Analysis of the {Bitcoin} Transaction Network}.
\newblock \bibinfo{journal}{\emph{PLoS ONE}} \bibinfo{volume}{9}, \bibinfo{number}{2} (\bibinfo{year}{2014}), \bibinfo{pages}{e86197}.
\newblock


\bibitem[Kondor et~al\mbox{.}(2021)]%
        {KondorDataset2021}
\bibfield{author}{\bibinfo{person}{D{\'a}niel Kondor}, \bibinfo{person}{M{\'a}rton P{\'o}sfai}, \bibinfo{person}{Istv{\'a}n Csabai}, {and} \bibinfo{person}{G{\'a}bor Vattay}.} \bibinfo{year}{2021}\natexlab{}.
\newblock \bibinfo{booktitle}{\emph{Bitcoin Transaction Network}}.
\newblock
\shownote{Version 2020, transactions up to February 2020}.
\newblock
\href{https://doi.org/10.5061/dryad.qz612jmcf}{doi:\nolinkurl{10.5061/dryad.qz612jmcf}}


\bibitem[Lin et~al\mbox{.}(2026)]%
        {LinEtAl2026}
\bibfield{author}{\bibinfo{person}{Ziqian Lin}, \bibinfo{person}{Qining Luo}, \bibinfo{person}{Dongze Wu}, \bibinfo{person}{Jie Shen}, \bibinfo{person}{Lingying Li}, \bibinfo{person}{Xunyi Nong}, {and} \bibinfo{person}{Zhenkai Qin}.} \bibinfo{year}{2026}\natexlab{}.
\newblock \showarticletitle{Detecting Illicit Transactions in {Bitcoin}: A Wavelet-Temporal Graph Transformer Approach for Anti-Money Laundering ({ChronoWave-GNN})}.
\newblock \bibinfo{journal}{\emph{Scientific Reports}} \bibinfo{volume}{16}, \bibinfo{number}{1} (\bibinfo{year}{2026}), \bibinfo{pages}{1548}.
\newblock
\href{https://doi.org/10.1038/s41598-025-23901-3}{doi:\nolinkurl{10.1038/s41598-025-23901-3}}


\bibitem[Lo et~al\mbox{.}(2023)]%
        {LoEtAl2023}
\bibfield{author}{\bibinfo{person}{Wai~Weng Lo}, \bibinfo{person}{Gayan~K. Kulatilleke}, \bibinfo{person}{Mohanad Sarhan}, \bibinfo{person}{Siamak Layeghy}, {and} \bibinfo{person}{Marius Portmann}.} \bibinfo{year}{2023}\natexlab{}.
\newblock \showarticletitle{Inspection-{L}: Self-Supervised {GNN} Node Embeddings for Money Laundering Detection in {Bitcoin}}.
\newblock \bibinfo{journal}{\emph{Applied Intelligence}}  \bibinfo{volume}{53} (\bibinfo{year}{2023}), \bibinfo{pages}{19406--19417}.
\newblock


\bibitem[Maganti(2026)]%
        {Maganti2026}
\bibfield{author}{\bibinfo{person}{Saket Maganti}.} \bibinfo{year}{2026}\natexlab{}.
\newblock \showarticletitle{When Graph Structure Becomes a Liability: A Critical Re-Evaluation of Graph Neural Networks for Bitcoin Fraud Detection under Temporal Distribution Shift}.
\newblock \bibinfo{journal}{\emph{arXiv preprint arXiv:2604.19514}} (\bibinfo{year}{2026}).
\newblock


\bibitem[Meiklejohn et~al\mbox{.}(2013)]%
        {Meiklejohn2013}
\bibfield{author}{\bibinfo{person}{Sarah Meiklejohn}, \bibinfo{person}{Marjori Pomarole}, \bibinfo{person}{Grant Jordan}, \bibinfo{person}{Kirill Levchenko}, \bibinfo{person}{Damon McCoy}, \bibinfo{person}{Geoffrey~M. Voelker}, {and} \bibinfo{person}{Stefan Savage}.} \bibinfo{year}{2013}\natexlab{}.
\newblock \showarticletitle{A Fistful of Bitcoins: Characterizing Payments Among Men with No Names}. In \bibinfo{booktitle}{\emph{Proceedings of the 2013 Internet Measurement Conference}} \emph{(\bibinfo{series}{IMC '13})}. \bibinfo{publisher}{ACM}, \bibinfo{address}{Barcelona, Spain}, \bibinfo{pages}{127--140}.
\newblock
\href{https://doi.org/10.1145/2504730.2504747}{doi:\nolinkurl{10.1145/2504730.2504747}}


\bibitem[Nguyen et~al\mbox{.}(2026)]%
        {NguyenBoersmaAcar2025}
\bibfield{author}{\bibinfo{person}{Linh Nguyen}, \bibinfo{person}{Marcel Boersma}, {and} \bibinfo{person}{Erman Acar}.} \bibinfo{year}{2026}\natexlab{}.
\newblock \showarticletitle{Detecting Fraud in Financial Networks: A Semi-supervised GNN Approach with Granger-Causal Explanations}. In \bibinfo{booktitle}{\emph{Explainable Artificial Intelligence}}, \bibfield{editor}{\bibinfo{person}{Riccardo Guidotti}, \bibinfo{person}{Ute Schmid}, {and} \bibinfo{person}{Luca Longo}} (Eds.). \bibinfo{publisher}{Springer Nature Switzerland}, \bibinfo{address}{Cham}, \bibinfo{pages}{330--353}.
\newblock
\showISBNx{978-3-032-08330-2}


\bibitem[Nicholls et~al\mbox{.}(2023)]%
        {NichollsEtAl2023FraudLens}
\bibfield{author}{\bibinfo{person}{Jack Nicholls}, \bibinfo{person}{Aditya Kuppa}, {and} \bibinfo{person}{Nhien-An Le-Khac}.} \bibinfo{year}{2023}\natexlab{}.
\newblock \showarticletitle{FraudLens: Graph Structural Learning for Bitcoin Illicit Activity Identification}. In \bibinfo{booktitle}{\emph{Proceedings of the 39th Annual Computer Security Applications Conference}} (Austin, TX, USA) \emph{(\bibinfo{series}{ACSAC '23})}. \bibinfo{publisher}{Association for Computing Machinery}, \bibinfo{address}{New York, NY, USA}, \bibinfo{pages}{324–336}.
\newblock
\showISBNx{9798400708862}
\href{https://doi.org/10.1145/3627106.3627200}{doi:\nolinkurl{10.1145/3627106.3627200}}


\bibitem[Paquet-Clouston et~al\mbox{.}(2019)]%
        {PaquetClouston2019}
\bibfield{author}{\bibinfo{person}{Masarah Paquet-Clouston}, \bibinfo{person}{Bernhard Haslhofer}, {and} \bibinfo{person}{Beno{\^\i}t Dupont}.} \bibinfo{year}{2019}\natexlab{}.
\newblock \showarticletitle{Ransomware Payments in the {Bitcoin} Ecosystem}.
\newblock \bibinfo{journal}{\emph{Journal of Cybersecurity}} \bibinfo{volume}{5}, \bibinfo{number}{1} (\bibinfo{year}{2019}), \bibinfo{pages}{tyz003}.
\newblock
\href{https://doi.org/10.1093/cybsec/tyz003}{doi:\nolinkurl{10.1093/cybsec/tyz003}}


\bibitem[Pareja et~al\mbox{.}(2020)]%
        {ParejaEtAl2020}
\bibfield{author}{\bibinfo{person}{Aldo Pareja}, \bibinfo{person}{Giacomo Domeniconi}, \bibinfo{person}{Jie Chen}, \bibinfo{person}{Tengfei Ma}, \bibinfo{person}{Toyotaro Suzumura}, \bibinfo{person}{Hiroki Kanezashi}, \bibinfo{person}{Tim Kaler}, \bibinfo{person}{Tao Schardl}, {and} \bibinfo{person}{Charles Leiserson}.} \bibinfo{year}{2020}\natexlab{}.
\newblock \showarticletitle{EvolveGCN: Evolving graph convolutional networks for dynamic graphs}. In \bibinfo{booktitle}{\emph{Proceedings of the AAAI conference on artificial intelligence}}, Vol.~\bibinfo{volume}{34}. \bibinfo{pages}{5363--5370}.
\newblock


\bibitem[Pendlebury et~al\mbox{.}(2019)]%
        {Pendlebury2019}
\bibfield{author}{\bibinfo{person}{Feargus Pendlebury}, \bibinfo{person}{Fabio Pierazzi}, \bibinfo{person}{Roberto Jordaney}, \bibinfo{person}{Johannes Kinder}, {and} \bibinfo{person}{Lorenzo Cavallaro}.} \bibinfo{year}{2019}\natexlab{}.
\newblock \showarticletitle{{TESSERACT}: Eliminating Experimental Bias in Malware Classification across Space and Time}. In \bibinfo{booktitle}{\emph{28th USENIX Security Symposium (USENIX Security 19)}}. \bibinfo{publisher}{USENIX Association}, \bibinfo{address}{Santa Clara, CA}, \bibinfo{pages}{729--746}.
\newblock
\showISBNx{978-1-939133-06-9}
\urldef\tempurl%
\url{https://www.usenix.org/conference/usenixsecurity19/presentation/pendlebury}
\showURL{%
\tempurl}


\bibitem[P{\'e}rez-Cano and Jurado(2025)]%
        {PerezCanoJurado2025}
\bibfield{author}{\bibinfo{person}{V{\'i}ctor P{\'e}rez-Cano} {and} \bibinfo{person}{Francisco Jurado}.} \bibinfo{year}{2025}\natexlab{}.
\newblock \showarticletitle{Fraud Detection in Cryptocurrency Networks---An Exploration Using Anomaly Detection and Heterogeneous Graph Transformers}.
\newblock \bibinfo{journal}{\emph{Future Internet}} \bibinfo{volume}{17}, \bibinfo{number}{1} (\bibinfo{year}{2025}), \bibinfo{pages}{44}.
\newblock


\bibitem[Raasveldt and M{\"u}hleisen(2019)]%
        {Raasveldt2019DuckDB}
\bibfield{author}{\bibinfo{person}{Mark Raasveldt} {and} \bibinfo{person}{Hannes M{\"u}hleisen}.} \bibinfo{year}{2019}\natexlab{}.
\newblock \showarticletitle{DuckDB: an Embeddable Analytical Database}. In \bibinfo{booktitle}{\emph{Proceedings of the 2019 International Conference on Management of Data}} (Amsterdam, Netherlands) \emph{(\bibinfo{series}{SIGMOD '19})}. \bibinfo{publisher}{Association for Computing Machinery}, \bibinfo{address}{New York, NY, USA}, \bibinfo{pages}{1981–1984}.
\newblock
\showISBNx{9781450356435}
\href{https://doi.org/10.1145/3299869.3320212}{doi:\nolinkurl{10.1145/3299869.3320212}}


\bibitem[Ron and Shamir(2013)]%
        {RonShamir2013}
\bibfield{author}{\bibinfo{person}{Dorit Ron} {and} \bibinfo{person}{Adi Shamir}.} \bibinfo{year}{2013}\natexlab{}.
\newblock \showarticletitle{Quantitative Analysis of the Full Bitcoin Transaction Graph}. In \bibinfo{booktitle}{\emph{Financial Cryptography and Data Security}}, \bibfield{editor}{\bibinfo{person}{Ahmad-Reza Sadeghi}} (Ed.). \bibinfo{publisher}{Springer Berlin Heidelberg}, \bibinfo{address}{Berlin, Heidelberg}, \bibinfo{pages}{6--24}.
\newblock
\showISBNx{978-3-642-39884-1}


\bibitem[{\v{S}}af{\'a}{\v{r}} et~al\mbox{.}(2026)]%
        {Safar2026}
\bibfield{author}{\bibinfo{person}{Miroslav {\v{S}}af{\'a}{\v{r}}}, \bibinfo{person}{Jan Pluskal}, \bibinfo{person}{Vladim{\'i}r Vesel{\'y}}, {and} \bibinfo{person}{Ond{\v{r}}ej Ry{\v{s}}av{\'y}}.} \bibinfo{year}{2026}\natexlab{}.
\newblock \showarticletitle{The Enemy of Reproducibility is Opacity: What's Inside the {Elliptic} {Bitcoin} Dataset (and Why It Is Wrong)}.
\newblock \bibinfo{journal}{\emph{Forensic Science International: Digital Investigation}}  \bibinfo{volume}{57} (\bibinfo{year}{2026}), \bibinfo{pages}{302124}.
\newblock
\shownote{Proceedings of DFRWS USA 2026}.
\newblock
\href{https://doi.org/10.1016/j.fsidi.2026.302124}{doi:\nolinkurl{10.1016/j.fsidi.2026.302124}}


\bibitem[Shaizat and Mussiraliyeva(2025)]%
        {ShaizatMussiraliyeva2025}
\bibfield{author}{\bibinfo{person}{Medet Shaizat} {and} \bibinfo{person}{Shynar Mussiraliyeva}.} \bibinfo{year}{2025}\natexlab{}.
\newblock \showarticletitle{Enhanced Identification of Illicit {Bitcoin} Transactions through Genetic Algorithm-Based Feature Selection}.
\newblock \bibinfo{journal}{\emph{Eastern-European Journal of Enterprise Technologies}} \bibinfo{volume}{4}, \bibinfo{number}{9 (136)} (\bibinfo{year}{2025}), \bibinfo{pages}{34--42}.
\newblock


\bibitem[Song and Gu(2023)]%
        {SongGu2023}
\bibfield{author}{\bibinfo{person}{Jialin Song} {and} \bibinfo{person}{Yijun Gu}.} \bibinfo{year}{2023}\natexlab{}.
\newblock \showarticletitle{HBTBD: A Heterogeneous Bitcoin Transaction Behavior Dataset for Anti-Money Laundering}.
\newblock \bibinfo{journal}{\emph{Applied Sciences}} \bibinfo{volume}{13}, \bibinfo{number}{15} (\bibinfo{year}{2023}).
\newblock
\showISSN{2076-3417}
\href{https://doi.org/10.3390/app13158766}{doi:\nolinkurl{10.3390/app13158766}}


\bibitem[Song et~al\mbox{.}(2025)]%
        {Song2025}
\bibfield{author}{\bibinfo{person}{Jie Song}, \bibinfo{person}{Sijia Zhang}, \bibinfo{person}{Pengyi Zhang}, \bibinfo{person}{Junghoon Park}, \bibinfo{person}{Yu Gu}, {and} \bibinfo{person}{Ge Yu}.} \bibinfo{year}{2025}\natexlab{}.
\newblock \showarticletitle{Illicit Social Accounts? Anti-Money Laundering for Transactional Blockchains}.
\newblock \bibinfo{journal}{\emph{IEEE Transactions on Information Forensics and Security}}  \bibinfo{volume}{20} (\bibinfo{year}{2025}), \bibinfo{pages}{391--404}.
\newblock
\urldef\tempurl%
\url{https://api.semanticscholar.org/CorpusID:274794309}
\showURL{%
\tempurl}


\bibitem[Trimborn et~al\mbox{.}(2024)]%
        {Trimborn2024}
\bibfield{author}{\bibinfo{person}{Simon Trimborn}, \bibinfo{person}{Hanqiu Peng}, {and} \bibinfo{person}{Ying Chen}.} \bibinfo{year}{2024}\natexlab{}.
\newblock \showarticletitle{Influencer detection meets network autoregression—Influential regions in the bitcoin blockchain}.
\newblock \bibinfo{journal}{\emph{Journal of Empirical Finance}}  \bibinfo{volume}{78} (\bibinfo{year}{2024}), \bibinfo{pages}{101529}.
\newblock


\bibitem[{U.S. Department of the Treasury}(2018)]%
        {OFAC2018}
\bibfield{author}{\bibinfo{person}{{U.S. Department of the Treasury}}.} \bibinfo{year}{2018}\natexlab{}.
\newblock \bibinfo{title}{Treasury Designates {Iran}-Based Individuals Who Helped Exchange Digital Currency Ransom Payments into {Iranian} Rial}.
\newblock \bibinfo{howpublished}{Press release}.
\newblock
\shownote{\url{https://home.treasury.gov/news/press-releases/sm556}}.
\newblock


\bibitem[Veli{\v{c}}kovi{\'c} et~al\mbox{.}(2017)]%
        {VelickovicEtAl2018}
\bibfield{author}{\bibinfo{person}{Petar Veli{\v{c}}kovi{\'c}}, \bibinfo{person}{Guillem Cucurull}, \bibinfo{person}{Arantxa Casanova}, \bibinfo{person}{Adriana Romero}, \bibinfo{person}{Pietro Lio}, {and} \bibinfo{person}{Yoshua Bengio}.} \bibinfo{year}{2017}\natexlab{}.
\newblock \showarticletitle{Graph attention networks}.
\newblock \bibinfo{journal}{\emph{arXiv preprint arXiv:1710.10903}} (\bibinfo{year}{2017}).
\newblock


\bibitem[Weber et~al\mbox{.}(2019)]%
        {WeberEtAl2019}
\bibfield{author}{\bibinfo{person}{Mark Weber}, \bibinfo{person}{Giacomo Domeniconi}, \bibinfo{person}{Jie Chen}, \bibinfo{person}{Daniel Karl~I Weidele}, \bibinfo{person}{Claudio Bellei}, \bibinfo{person}{Tom Robinson}, {and} \bibinfo{person}{Charles~E Leiserson}.} \bibinfo{year}{2019}\natexlab{}.
\newblock \showarticletitle{Anti-money laundering in bitcoin: Experimenting with graph convolutional networks for financial forensics}.
\newblock \bibinfo{journal}{\emph{arXiv preprint arXiv:1908.02591}} (\bibinfo{year}{2019}).
\newblock


\end{thebibliography}

\newpage
\appendix

\section{Feature definitions}
\label{app:ael55-defs}

\Cref{tab:ael55-defs} gives the computation of each of the 55 features. In the leakage-resistant set, the transactions of a training address are restricted to blocks up to the cutoff, block 457{,}734.

\begin{table}[h]
\caption{Computation of the 55 features of an address $a$.}
\label{tab:ael55-defs}
\footnotesize
\begin{tabular}{@{}l l@{}}
\toprule
Feature & Computation \\
\midrule
\multicolumn{2}{@{}l}{\textit{Per-transaction quantities}, for $t \in T$} \\
Net amount sent      & $s_t = \max(\mathrm{in}_t - \mathrm{out}_t,\, 0)$ \\
Net amount received  & $r_t = \max(\mathrm{out}_t - \mathrm{in}_t,\, 0)$ \\
Amount transacted    & $v_t = s_t + r_t$ \\
Fee                  & $f_t = \max(I_t - O_t,\, 0)$ \\
Fee share            & $\phi_t = f_t \cdot \max(\mathrm{in}_t, \mathrm{out}_t) / I_t$ \\
\addlinespace
\multicolumn{2}{@{}l}{\textit{Volumes} (15)} \\
Amount transacted (5)   & $\operatorname{stat}_{t \in T}(v_t)$ \\
Net amount sent (5)     & $\operatorname{stat}_{t \in T}(s_t)$ \\
Net amount received (5) & $\operatorname{stat}_{t \in T}(r_t)$ \\
\addlinespace
\multicolumn{2}{@{}l}{\textit{Fees} (10)} \\
Fee (5)       & $\operatorname{stat}_{t \in T}(f_t)$ \\
Fee share (5) & $\operatorname{stat}_{t \in T}(\phi_t)$ \\
\addlinespace
\multicolumn{2}{@{}l}{\textit{Block gaps} (15)} \\
All transactions (5)       & $\operatorname{stat}\, \Delta(T)$ \\
Sending transactions (5)   & $\operatorname{stat}\, \Delta(S)$ \\
Receiving transactions (5) & $\operatorname{stat}\, \Delta(R)$ \\
\addlinespace
\multicolumn{2}{@{}l}{\textit{Counterparties} (6)} \\
Interactions per counterparty (5) & $\operatorname{stat}_{c \in C}(n_c)$ \\
Repeated counterparties (1)       & $\lvert \{ c \in C \mid n_c > 1 \} \rvert$ \\
\addlinespace
\multicolumn{2}{@{}l}{\textit{Transaction counts} (3)} \\
Transactions (1)           & $\lvert T \rvert$ \\
Sending transactions (1)   & $\lvert S \rvert$ \\
Receiving transactions (1) & $\lvert R \rvert$ \\
\addlinespace
\multicolumn{2}{@{}l}{\textit{Lifetime} (6)} \\
First block (1)           & $\min_{t \in T} b_t$ \\
Last block (1)            & $\max_{t \in T} b_t$ \\
Lifetime (1)              & $\max_{t \in T} b_t - \min_{t \in T} b_t$ \\
First sending block (1)   & $\min_{t \in S} b_t$, or 0 if $S$ is empty \\
First receiving block (1) & $\min_{t \in R} b_t$, or 0 if $R$ is empty \\
Time steps (1)            & number of distinct Elliptic time steps of $t \in T$ \\
\bottomrule
\end{tabular}

\smallskip
\begin{minipage}{\linewidth}\footnotesize \textit{Note:} $t$ denotes a transaction and $c$ a counterparty address. $T$ is the set of Elliptic transactions in which $a$ appears as an input or an output, and $S = \{t \in T \mid s_t > 0\}$ and $R = \{t \in T \mid r_t > 0\}$ are its sending and receiving transactions. For $t \in T$, $\mathrm{in}_t$ and $\mathrm{out}_t$ are the amounts $a$ contributes as inputs and collects as outputs, $I_t$ and $O_t$ are the total input and output of $t$, and $b_t$ is its block. Amounts are in bitcoin, and $\phi_t = 0$ if $I_t = 0$. $\operatorname{stat}$ stands for the total, minimum, maximum, mean, and median, so each row containing it gives five features. The number in parentheses is the number of features in a row. $\Delta(X)$ lists the differences between the blocks of consecutive transactions of $X$ ordered by block, and is $(0)$ if $X$ has fewer than two transactions. The counterparties of $a$ in $t$ are the output addresses of $t$ if $t \in S$ and its input addresses if $t \in R$, excluding $a$. $C$ is the set of all counterparties of $a$, and $n_c$ is the number of transactions in which $c$ is a counterparty of $a$. $\lvert X \rvert$ denotes the number of elements of a set $X$.
\end{minipage}
\end{table}

\section{Validation of the recovered features}
\label{app:validation}

The one-percent check of \Cref{subsec:reproduction} combines a relative tolerance of 0.01 with an absolute tolerance of $10^{-4}$ bitcoin (10{,}000 satoshis), about one twentieth of the median transaction fee of the labelled addresses (0.00185 bitcoin). The absolute term keeps differences in very small amounts from counting as divergences. Under this rule, 86 addresses (0.03\%) differ from the released values by more than one percent on at least one of the 25 value features. Replacing the absolute term with one satoshi raises the number to 137. The 51 additional addresses differ by at most 0.0003 bitcoin. Most differ only on the fee share, whose values for these addresses are of the order of $10^{-5}$ bitcoin, and 15 differ on every volume feature by the same 0.000027 bitcoin where the released value is zero. Under the stricter rule, the per-transaction fee share matches the released values for 99.96\% of addresses, against 6.00\% for the single ratio of total fees to total amount transacted. Dropping either set of divergent addresses leaves model performance unchanged (\Cref{tab:divergent-sensitivity}).

\begin{table}[h]
\caption{Sensitivity of the reproduction to the divergent addresses.}
\label{tab:divergent-sensitivity}
\resizebox{\columnwidth}{!}{%
\begin{tabular}{l ccc ccc ccc}
\toprule
& \multicolumn{3}{c}{All addresses} & \multicolumn{3}{c}{86 dropped} & \multicolumn{3}{c}{137 dropped} \\
\cmidrule(lr){2-4}\cmidrule(lr){5-7}\cmidrule(lr){8-10}
Model & P & R & F1 & P & R & F1 & P & R & F1 \\
\midrule
LR & 0.492 & 0.062 & 0.110 & 0.481 & 0.065 & 0.114 & 0.483 & 0.064 & 0.114 \\
RF & \sd{0.911}{0.004} & \sd{0.786}{0.003} & \sd{0.844}{0.002}
   & \sd{0.915}{0.004} & \sd{0.786}{0.002} & \sd{0.846}{0.002}
   & \sd{0.912}{0.005} & \sd{0.785}{0.003} & \sd{0.844}{0.004} \\
\bottomrule
\end{tabular}}

\smallskip
\begin{minipage}{\linewidth}\footnotesize
\textit{Note:} LR, logistic regression; RF, random forest with 50 trees, mean over 10 seeds with the standard deviation below. Recovered lifetime features, positional 70/30 split, and the pipeline of \citet{ElmougyLiu2023} as in \Cref{tab:reproduction}. The 86 addresses diverge from the released values under an absolute tolerance of $10^{-4}$ bitcoin, the 137 under one satoshi, both with a relative tolerance of 0.01. Precision, recall, and F1 at the default 0.5 threshold.
\end{minipage}
\end{table}

\end{document}